\documentclass[10pt,twoside,BCOR7mm,DIV15,headinclude,footexclude,cleardoubleempty,idxtotoc]{scrartcl}

\usepackage[english]{babel}
\usepackage{graphicx}
\usepackage{hyperref}
\usepackage{scrpage2}
\usepackage{hyperref}
\usepackage{ifthen}

\makeatletter
\renewcommand{\@biblabel}[1]{}
\renewcommand{\@cite}[2]{%
{#1\ifthenelse{\boolean{@tempswa}}{,#2}{}}}
\makeatother

\hypersetup{breaklinks=true
,colorlinks=true,linkcolor=black,urlcolor=blue
,citecolor=black}

\ofoot{\thepage}
\ifoot{}

\setheadsepline{1pt}

\setkomafont{pagehead}{\normalfont\sffamily}
\setkomafont{pagenumber}{\normalfont\rmfamily}

\usepackage{booktabs}
\usepackage{amsmath}
\usepackage{amssymb}
\usepackage{multicol}
\usepackage{float}

\makeatletter
\newcommand{\listofcontributions}{\@starttoc{con}}

\newcommand{\l@contribution} {\@dottedtocline{1}{1.5em}{2.3em}}
\makeatother

\newenvironment{contribution}{
\setcounter{section}{0}
\setcounter{figure}{0}
\setcounter{table}{0}
\begin{flushleft}
{\em Clumping in Hot Star Winds \\
W.-R.\ Hamann, A.\ Feldmeier \& L.\ Oskinova, eds.\\
Potsdam: Univ.-Verl., 2007 \\
URN: http://nbn-resolving.de/urn:nbn:de:kobv:517-opus-13981
} 
\end{flushleft}
}{
\newpage
\lehead{}
\rohead{}
}

\begin{document}

\setlength{\baselineskip}{2.5ex}

\begin{contribution}

\lehead{J.\ Krti\v cka, J.\ Puls, \& J.\ Kub\'at}

\rohead{The influence of clumping on predicted O star wind parameters}

\begin{center}
{\LARGE \bf The influence of clumping on predicted O star wind parameters}\\
\medskip

{\it\bf J.\ Krti\v cka$^1$, J.\ Puls$^2$ \& J.\ Kub\'at$^3$}\\

{\it $^1$Masarykova universita, Brno, Czech Republic}\\
{\it $^2$Universit\"atssternwarte M\"unchen, M\"unchen, Germany}\\
{\it $^3$Astronomick\'y \'ustav, Ond\v{r}ejov, Czech Republic}

\begin{abstract}
We study the influence of clumping on the predicted wind structure of O-type
stars. For this purpose we artificially include clumping into our stationary
wind models. When the clumps are assumed to be optically thin, the radiative
line force increases compared to corresponding unclumped models, with a
similar effect on either the mass-loss rate or the terminal velocity
(depending on the onset of clumping). Optically thick clumps, alternatively,
might be able to decrease the radiative force. 
\end{abstract}
\end{center}

\begin{multicols}{2}

\section{Introduction}

Theoretical models of hot star winds led to the conclusion that mass-loss has
a significant impact on the stellar evolution in the upper HRD. This
conclusion was supported by a relatively good agreement between these
models and observational data. 

However, both theoretical predictions and observational values were derived,
in a first approximation, by assuming the stellar wind to be a smooth,
spherically symmetric outflow. This picture might be
inadequate due to the existence of a strong instability related to
radiative line driving 
(see Feldmeier, this volume, and references therein). The influence of
corresponding spatially inhomogeneous wind structures (``clumps'') on the
basic wind properties (i.e., mass-loss rate and velocity field) and the
emergent spectrum was believed to be insignificant.

This changed with the application of NLTE models that were able to account for
the influence of clumping on hot star wind spectra. The reason that the
effect of clumping is not immediately apparent in the wind spectra is that
most diagnostical features depend on the {\it product} $\sqrt
{C_\text{c}}\dot M$, where the "clumping factor" ${C_\text{c}}\geq1$
relates the density inside the clump $\rho^+$ with the mean wind density
$\langle\rho\rangle$,
\begin{equation}
\label{krticka:cc}
\rho^+=C_\text{c}\langle\rho\rangle.
\end{equation}
Consequently, spectra from winds with a large clumping factor but small
mass-loss rate can mimic those from winds with weak clumping but large
mass-loss rate. If ${C_\text{c}} > 1$, then the
mass-loss rates derived from such diagnostics are overestimated by a factor
of $\sqrt{C_\text{c}}$. Fortunately, some spectral properties may be used to
break this degeneracy, and to ``observationally'' 
estimate the value of ${C_\text{c}}$.
For example, 
Martins et al.~(\cite{krticka:martclump})
derived clumping factors of 
$10$ and $100$ for studied
Galactic O-stars, implying
a decrease of the estimated mass-loss rate by factors of 
$3$ and $10$
(see also Bouret, this volume, and Puls et al., this volume, and references
therein).

Thus, if the actual mass-loss rates of hot stars are really lower
than assumed before, then there is a significant discrepancy between the
observations and theory.
The reason for this discrepancy is not yet known. A possibility might be 
that clumping may shift the ionization balance in such a way that the final
radiative force is lower, resulting in lower mass-loss rates.  To test this
possibility we have included clumping into our wind models and studied its influence
on the basic wind properties of hot stars.

\section{Model stars}

The wind models are calculated for O-type stellar parameters 
based on the recent
calibrations by Martins et al.~(\cite{krticka:okali},
see also Fig.\,\ref{krticka:hrdhm}).

\begin{figure}[H]
\begin{center}
\includegraphics[width=0.9\columnwidth]{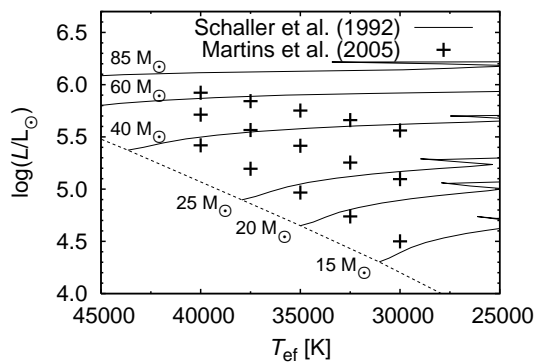}
\caption{Parameters of studied stars in the HR diagram. Overplotted are the
evolutionary tracks from Schaller et al.~(\cite{krticka:salek}).%
\label{krticka:hrdhm}}
\end{center}
\end{figure}

\section{Wind models}

For our study we used the spherically symmetric, stationary wind models
developed by Krti\v{c}ka \& Kub\'{a}t~(\cite{krticka:nltei}). These models solve the
equations of statistical equilibrium together with the equations of 
radiative transfer. The calculated occupation numbers are used to derive the
radiative force (in the Sobolev approximation) and the radiative
heating/cooling terms. This enables us to obtain the radial stratification 
of velocity, density and temperature in the wind and finally to predict the
wind mass-loss rate $\dot M$ and terminal velocity $v_\infty$. For the 
studied O-stars, $\dot M$
derived from our models
depends on the stellar luminosity $L$ and the
effective temperature $T_\text{eff}$ on average as
\begin{eqnarray}
\left(\!{\frac{\dot M}{1\,\text{M}_\odot\,\text{year}^{-1}}}\!\right)\!=8.13\times10^{-7}
\left(\!{\frac{L_*}{3\times10^{5}\,\text{L}_\odot}}\!\right)^{2.05}\!\!\!\\\times
\left(\!{\frac{T_\text{eff}}{3.5\times10^4\,\text{K}}}\!\right)^{3.78}\nonumber.
\end{eqnarray}
The advantage of our models is the self-consistent solution of the momentum
equation, though the radiative transfer is treated in a simplified way.

\section{Optically thin clumps}

The assumption of optically thin clumps is widely used for studying the 
influence of clumping on the wind spectra (e.g.,
Martins et al.~\cite{krticka:martclump},
Puls et al, this volume).

To include optically thin clumps into our models we
modified the equations of statistical equilibrium, by using an electron
density 
$\rho_\text{e}^+={C_\text{c}}\langle{\rho_\text{e}}\rangle$,
opacity $\langle\chi\rangle={\chi^+}/{{C_\text{c}}}$, and
emissivity $\langle\eta\rangle={\eta^+}/{{C_\text{c}}}$.
The superscript $+$ denotes values inside the (homogeneous) clumps
and the quantities inside brackets corresponding volume averages.

\subsection{Influence of clumping}

To investigate the influence of clumping on the stellar wind we have 
calculated a wind model of an O-type giant at
$T_\text{eff}=35\,000\,\text{K}$, assuming the wind to be smooth
($C_\text{c}=1$) close to the star ($ r<2R_{*}$), and to be clumped
($C_\text{c}=10$) in the outer regions ($r>2R_{*}$, see
Figs.~\ref{krticka:350_3_ion}, \ref{krticka:350_3_r2b}).

The presence of clumping leads to an increase of the electron density inside
the clumps. Consequently, the recombination rates become higher and the wind
less ionized (see Fig.~\ref{krticka:350_3_ion}). Since lower ions are able
to accelerate the stellar wind more efficiently than the higher ones (due to
a larger number of driving lines), the
radiative force increases, which, in our case, leads to an increase in 
wind velocity (see Fig.~\ref{krticka:350_3_r2b}).

\begin{figure}[H]
\begin{center}
\includegraphics[width=0.9\columnwidth]{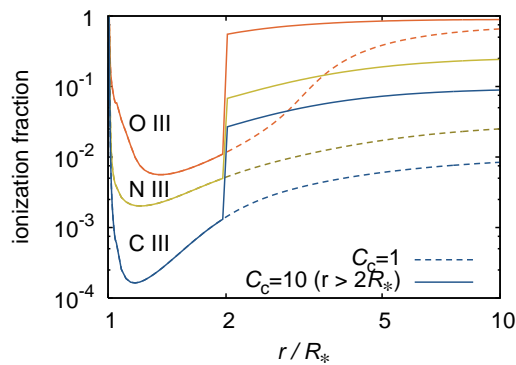}
\caption{Influence of clumping on the ionization fractions of selected ions.%
\label{krticka:350_3_ion}}
\end{center}
\end{figure}

\begin{figure}[H]
\begin{center}
\includegraphics[width=0.9\columnwidth]{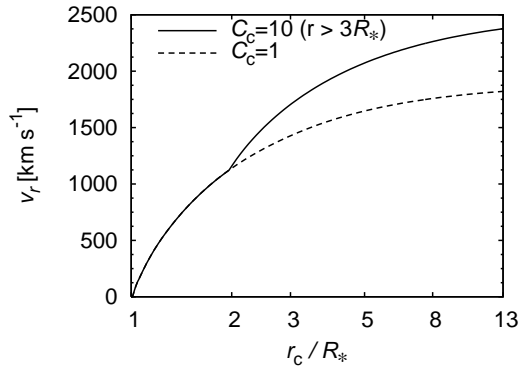}
\caption{Influence of clumping on the wind velocity.\label{krticka:350_3_r2b}}
\end{center}
\end{figure}

\subsection{Radially constant clumping factor}

The influence of clumping on the wind parameters depends on the radial onset
of clumping. If clumping starts above the critical point (below which the
mass-loss rate is determined), then the terminal velocity increases (see
Fig.~\ref{krticka:350_3_r2b}). On the other hand, if clumping starts below
the critical point, then the wind mass-loss rate becomes larger.
\begin{table}[H]
\begin{center}
\caption{Average increase of $\dot M$ for constant $C_\text{c}$}
\label{krticka:dmdtchuchprum}
\medskip
\begin{tabular}{cccccc}
\toprule
$C_\text{c}$ & 1& $3.16$ & $10$ & $31.6$ & $100$ \\
\midrule
$\dot M(C_\text{c})/\dot M(C_\text{c}=1)$ & 1& 1.48 & 2.15 & 3.17 & 4.57 \\
\bottomrule
\end{tabular}
\end{center}
\end{table}
In case of radially constant clumping, the mass-loss rate increases
significantly, and the predicted wind-momentum rate is much higher than for
a smooth wind at same parameters (see Fig.~\ref{krticka:mom_porov}).

\section{Clumps larger than $\boldsymbol L_\text{Sob}$}

Individual clumps may be larger than the Sobolev length $L_\text{Sob}$.
Assuming
the velocity gradient inside the clumps to be the same as in the corresponding smooth
wind, we account for clumps being larger than
$L_\text{Sob}$
by using the same modifications as in the case of optically thin clumps, but
additionally increasing the Sobolev optical depth by $C_\text{c}$, and
decreasing the radiative force per unit of volume by a factor of
$C_\text{c}$.
\begin{figure}[H]
\begin{center}
\includegraphics[width=0.85\columnwidth]{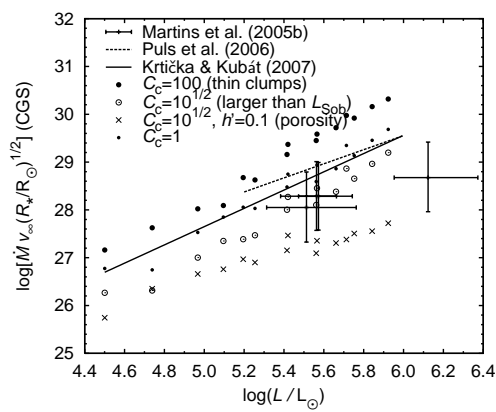}
\caption{Influence of wind inhomogeneities on the modified wind-momentum
rate.\label{krticka:mom_porov}}
\end{center}
\end{figure}
The inclusion of clumps larger than $L_\text{Sob}$ into our wind models
leads to a decrease of the radiative force. Thus, and for radially constant
$C_\text{c}$, the mass-loss rate decreases significantly
(Fig.~\ref{krticka:mom_porov}).

\section{Influence of porosity}

Wind porosity (Owocki et al.~\cite{krticka:owosha}) 
can be introduced into the wind models by additional decreasing the
continuum opacity, 
$\chi_\text{eff}=\frac{\langle\chi\rangle}{1+rh'\langle\chi\rangle}$, where
$h'r$ is the porosity length. The continuum emissivity has to be modified by
the same amount.

This effect leads to a significant increase of the wind ionization.
Thus, porosity leads
to
a decrease of the radiative force, and also the
mass-loss rate may decrease if the wind is porous below the critical point
(see Fig.~\ref{krticka:mom_porov} for the results for radially constant
$C_\text{c}$ and $h'$).

\section{Discussion}

Our results are in agreement with those derived from a similar investigation
conducted by de~Koter \& Muijres (this volume): Indeed, {\it wind
inhomogeneities influence the predicted wind parameters}. An approach which
roughly corresponds to the results of time-dependent simulations (i.e.,
clumps which are optically thin at most frequencies, ``starting'' above the
critical point) does not improve the agreement between theory and
observations. Clumps assumed to be larger than the Sobolev length (and 
starting below the critical point) may provide a better agreement
between theory and observations, both in terms of mass-loss rates
(Fig.~\ref{krticka:mom_porov}) and P~V ionization fractions
(Fig.~\ref{krticka:mion_P_5}). Note, however, that the approach chosen by us
strongly contrasts an important aspect of time-dependent simulations, namely
that the velocity gradient inside the clumps is predicted to be much
shallower than that of a corresponding smooth wind. This problem will be
considered in a follow-up investigation.
\begin{figure}[H]
\begin{center}
\includegraphics[width=0.9\columnwidth]{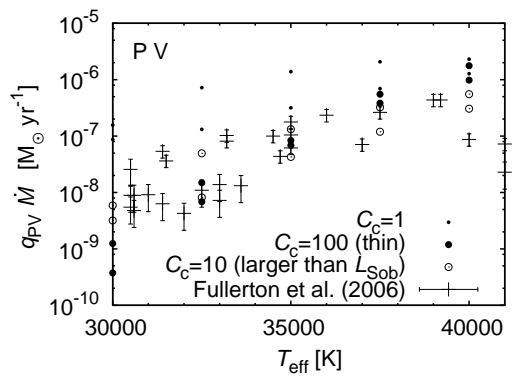}
\caption{Influence of different types of wind inhomogeneities on the P~V
ionization.\label{krticka:mion_P_5}}
\end{center}
\end{figure}

\paragraph*{Acknowledgements}

This research was supported by grants GA \v CR 205/07/0031 and GA AV \v{C}R
B301630501.


\end{multicols}

\end{contribution}


\end{document}